\documentclass[12pt]{iopart}
\usepackage{graphicx}
\begin{document}
\title{ The Landau-Zener Model with Decoherence:\\
 The case S=1/2}
\author{V G  Benza\dag and G Strini\ddag}
\address{\dag\ Facolt\`a di Scienze, Universit\`a dell'Insubria, sede di Como
Via Valleggio 11, Como, 
also I.N.F.M., Unit\`a di Como}
\address{\ddag\ Dipartimento di Fisica, Universit\`a di Milano 
Via Celoria 16, Milano}

\begin{abstract}
We study the dynamics of a spin coupled to an oscillating
magnetic field, in the presence of decoherence and dissipation.
In this context we solve the master equation for the Landau-Zener problem,
both in the unitary
and in the irreversible case.
We show that a single spin can be ``magnetized'' 
in the direction parallel to the oscillating bias. 
When  decay from upper to lower level is taken into account,
hysteretic behavior is obtained.
\end{abstract}

\maketitle

\section{Introduction}

This paper concerns the dynamics of a driven 
two-level system with  decoherence and dissipation.
Such a simple quantum-mechanical problem has been  studied
in various contexts, as, e.g., nuclear and atomic physics.
What follows is predominantly
motivated by molecular magnets
and quantum information, as long as
these topics suggest new perspectives to the old problem.
\par
The perspective of  a comprehensive 
theory  of molecular magnets stimulated a large amount of work.
Rather than giving an extensive review,
we merely quote few topics in this area.
A  1/2 spin (the central spin) has been introduced in modeling 
the macroscopic spin tunneling
(for a short review see, e.g., Ref.\cite{TB});
dissipative spin reversal, incoherent Zener 
tunneling, decoherence  were also considered 
 \cite{CWMBB1, CWMBB2, LELO, DKH, DS}.
\par
 Work more directly related to ours can be found in Refs. \cite{BBGW,
 BC, FMO}. 
In classifying the  behaviors of our system, we find effects 
such as single spin dynamical magnetization and hysteresis; 
the  meaning of these
terms will be made clear in the sequel.
\par
In the presence of an oscillating external field,
the evolution is characterized by a sequence of
level-crossings, where the Landau-Zener (L-Z) model 
\cite{LL,  Z} applies,
separated by intervals of ``normal'' evolution.
\par
We classify the dynamics from
the behavior in the L-Z regime and in the ``normal'' regime.
In order to do that, we first solve  the Landau-Zener
problem in the master equation for the density matrix.
This is done both in the unitary and in the  
irreversible case.

\par 
Our starting point is Kraus's \cite{K} representation theorem,
for the density matrix of a two-level system: a short reminder
of the theory is given in Sections 2,3;
for the sake of completeness, in  Appendix I we recall the  
conditions for the complete positivity of the evolution \cite{RSW}.

\par  
In Section 4 we discuss the unitary case
by means of  qualitative methods. 
We examine high frequency fields as well 
as quasistatic fields.
We show that a ``magnetized'' spin can be sustained, in the
direction of a zero-average oscillating bias.
\par 
The non-unitary case,
where dissipation and decoherence are taken into account,
is discussed in Section 5. 
We show that
  the L-Z problem  can be solved
 in the irreversible case;
the solution is explicit for two-level
systems.
\par
We find the lowest order (in the ${1/t}$ expansion) renormalizations 
of the dissipation and decoherence parameters:
  the non-unitary processes do not modify the scattering 
properties of the solution at the level crossings.
After  analyzing
 the  single-spin ``magnetization'' with  decoherence,
  we take into account  the  relaxation 
from the upper to the lower level.  In section 6 we show that
 an oscillating external field gives rise to  hysteretic behavior. 

\par

\section{ Master equation for the two-level system}

We start with the following master equation, for the density matrix
of a two-level system:

\begin{equation}
\label{RO}
{\dot \rho} \;=\; - \, \frac{i}{\hbar}\, [H,\, \rho (t)]\,+\,
\frac{1}{2}\, \sum_{i,j=1}^{3}\,
A_{i j} \, ( [\sigma_{i}, \, \rho (t) \, \sigma^{\dagger}_{j}] \,+\,
[\sigma_{i} \, \rho(t),\, \sigma_{j}^{\dagger}]) 
\end{equation}

\noindent
where $H$ is associated with  reversible dynamics,
the matrix ${\bf A}$ is 
Hermitean and the $\sigma_{i}$'s are the Pauli matrices. 

The Hamiltonian is parametrized by 3 real parameters:

\begin{equation}
H \;=\; \frac{1}{2} \{
\Delta \, \sigma_{1} \,+\,
\Delta' \, \sigma_{2} \,+\,
\omega_{0} \, \sigma_{3} 
\}
\end{equation}

\noindent
and further 9 real parameters label  the $3\times3$ hermitean matrix  
$A_{ij}$.
\par

Notice that this is consistent with the Kraus 
representation \cite{K},
 where the time evolution of a two-level system is 
characterized by
12 real parameters.
\par
As it is well known, the coefficients of  ${\bf A}$ are  
the correlation
functions of the macrosystem coupled with the spin.
\par 

We then write Eq. \ref{RO}
in the standard representation:

\begin{equation}
\rho (t) \;=\; \frac{1}{2}\,\left[
\begin{array}{cc}
1\, + \,Z & X \,-\, i Y\\
X \,+\, i Y & 1 \,-\, Z
\end{array}\right]
\hspace{.5 in}
{\bf v} \;=\; \left[
\begin{array}{c}
X \\
Y \\
Z
\end{array}\right]
\end{equation}

we have:

\begin{equation}
\label{V}
{\dot {\bf v}} (t) \;=\; M \, {\bf v} (t) \,+\, {\bf C}
\end{equation}

The unitary contribution in M  has the form:

\begin{equation}
\label{REV}
\left[
\begin{array}{c}
{\dot X}\\
{\dot Y}\\
{\dot Z}
\end{array}\right]_{H}
\;=\;\left[
\begin{array}{ccc}
0 & - \, \omega_{0} & \Delta' \\
\omega_{0} & 0 & - \, \Delta \\
- \, \Delta' & \Delta & 0
\end{array}\right]
\left[
\begin{array}{c}
 X \\
 Y \\
 Z
\end{array}\right]
\end{equation}

and the decoherence and dissipation processes give:

$$
\left[
\begin{array}{c}
{\dot X}\\
{\dot Y}\\
{\dot Z}
\end{array}\right]_{D}
=\left[
\begin{array}{ccc}
- \,2 \, (A_{22} + A_{33}) &
(A_{12} + A_{21}) &
 (A_{13} + A_{31}) \\
 (A_{12} + A_{21}) &
- 2  (A_{33} + A_{11}) &
 (A_{23} + A_{32}) \\
 (A_{13} + A_{31}) &
 (A_{23} + A_{32}) &
- 2  (A_{11} + A_{22}) 
\end{array}\right]
\left[
\begin{array}{c}
X \\
Y \\
Z
\end{array}\right]
+\left[
\begin{array}{c}
2i( A_{23} - A_{32}) \\
2i( A_{31} - A_{13}) \\
2i( A_{12} - A_{21}) \\
\end{array}\right]
$$

or:

$$
{\dot{\bf v}}_{D}\;=\; D \, {\bf v} \, \,+\, {\bf C}
$$

In Eq. \ref{V}, the explicit form of M is made more readable by
letting:

$$
\gamma_{1} \;=\; 2 \, (A_{22} \,+\, A_{33})
\hspace{.5 in}
\gamma_{2} \;=\; 2 \, (A_{33} \,+\, A_{11})
\hspace{.5 in}
\gamma_{3} \;=\; 2 \, (A_{11} \,+\, A_{22})
$$
$$
\alpha \;=\; (A_{12} \,+\, A_{21})
\hspace{.5 in}
\beta \;=\; (A_{13} \,+\, A_{31})
\hspace{.5 in}
\gamma \;=\; (A_{23} \,+\, A_{32})
$$
$$
C_{1} \;=\; 2 \,i \,  (A_{23} \,-\, A_{32})
\hspace{.5 in}
C_{2} \;=\; 2 \,i \,  (A_{31} \,-\, A_{13})
\hspace{.5 in}
C_{3} \;=\; 2 \,i \,  (A_{12} \,-\, A_{21})
$$
so that:

\begin{equation}
\left[
\begin{array}{c}
{\dot X}\\
{\dot Y}\\
{\dot Z}
\end{array}\right]
\;=\;\left[
\begin{array}{ccc}
-  \gamma_{1} & \alpha - \omega_{0} & \beta + \Delta' \\
\alpha + \omega_{0} & -  \gamma_{2}  & \gamma - \Delta \\
\beta - \Delta'  & \gamma + \Delta & -  \gamma_{3}
\end{array}\right]
\;\left[
\begin{array}{c}
X \\
Y \\
Z
\end{array}\right]
\,+\,\left[
\begin{array}{c}
C_{1} \\ C_{2} \\ C_{3}
\end{array}\right]
\end{equation}
\section{Geometrical interpretation of the 12 parameters}

The phase space of the system is a  Poincar\'e surface, undergoing
elementary instantaneous transformations, according with
 $M$ and ${\bf C}$.
\par
More specifically, while the antisymmetric part determines rotations,
the symmetric part gives dilatations along 3 principal axes.
The inhomogeneous term  translates the surface.

\par

The connection with  Kraus's  evolution of the density matrix
over a finite time interval  is readily verified in the 
case of constant coefficients; in such a case, by explicit time integration,
one gets the affine \cite{AL, K, S} \,map:

\begin{equation}
\label{KK}
\rho' \;=\; \sum_{i} \, \mathcal{ A}_{i} \, \rho \,
\mathcal{ A}_{i}^{\dagger}
\hspace{.5 in}
 \sum_{i} \, \mathcal{A}_{i}^{\dagger}  \,\mathcal{ A}_{i} \;=\; {\bf 1}
\end{equation}

\noindent
which again depends on 12 parameters.
\par
In our problem problem
the hamiltonian is time-dependent:
the Poincar\'e  surface still  undergoes an affine transformation
under a finite time evolution, but 
the operators $\mathcal{A}_{i}$ 
are fairly complicate functions of $M(t)$ and $C_{i}(t)$.

We  start by assuming
a time-independent
 matrix ${\bf A}$
in eq. \ref{RO}.
The eigenvectors of $D$ identify
the dilatation axes of the Poincar\'e surface.
The choice of  representation
for the hamiltonian part, given by  $M$,
fixes a second, time-dependent frame.
\par

Everything greatly simplifies if the two frames coincide,
i.e. if $D$ is diagonal:
$$
\alpha \;=\; \beta\;=\; \gamma \;=\; 0
$$
\par
The parameters are then reduced from 12 to 9: 
 3 of them come from  the hamiltonian, 3 from
decoherence and dissipation,
$(\gamma_{1}, \, \gamma_{2}, \, \gamma_{3})\,$, 
and the last  3 from the inhomogeneous terms $C_{i}$.

E.g., in the case of an isotropic thermal bath with average phonon
number ${\bar n}$ one has \cite{S}:
$$
\gamma_{1} \;=\; \gamma_{2} \;=\; - \, \gamma \, ( {\bar n} \,+\, \frac{1}{2} )
\hspace{.5 in}
\gamma_{3} \;=\; - \, \gamma \, ( 2 \, {\bar n} \,+\, 1)
$$
$$
C_{1} \;=\; C_{2} \;=\; 0
\hspace{.5 in}
C_{3} \; = \; - \, \gamma
$$
 
\section{Unitary evolution}

In this section we illustrate the dynamics in the unitary case.
The external field is taken, without loss of generality, in the x-z
plane: this implies $\Delta'=0$.
We consider an oscillating z-component: $B_{z}\,=\, \omega_{0}(t)$,
and a constant x-component: $B_{x}\,=\, \Delta$.
 
The evolution can  be depicted as a sequence 
of L-Z crossings
and ``normal'' regimes.
\par
We characterize the ``normal'' regime with the condition 
$|\omega_{0}(t)| \, \gg \Delta$:
it  is dominated by   rotations  
 in the plane $X,\,Y$, as one can verify by inspection of eq. \ref{REV}.
\par
One must not expect that the representative
on the Poincar\'e surface  retraces  its path backwards 
in coincidence with the external field: in fact the solution
is the time-ordered exponential of an operator.
This is  particularly true 
when  $\omega_{0}(t)$ is of the  order of  $\Delta$.
On the other hand, as long as the magnetic field
 $\{ \Delta,\,\omega_{0}(t)\}$ 
runs over a line in the plane 
 $\{ B_{x},\,B_{z}\}$, no Berry phase is to be expected.

We take, as an initial condition, the ``fully magnetized'' configuration 
$Z= \pm 1$.
We have then:

\begin{equation}
\label{Ham}
{\dot X}  \;=\; -\, \omega_{0} \, Y 
\hspace{.5 in}
{\dot Y}  \;=\;  \omega_{0} \, X \,-\, \Delta \, Z
\hspace{.5 in}
{\dot Z} \;=\; 
 \Delta \, Y
\end{equation}

$$
\omega_{0} \;=\; B_{0} \, cos ( \Omega_{0} \, t)
$$

\noindent
where $\Omega_{0}$ \ is the bias frequency.

\par
The Schr\"odinger equation corresponding to system \ref{Ham}
has the form:

\begin{eqnarray}
\label{Schrod}
i  {\dot a_{1}} &=& {1 \over 2} ( \omega_{0}  a_{1} +  \Delta  a_{2})\\
i  {\dot a_{2}} &=& {1 \over 2} (- \omega_{0}  a_{2} + \Delta  a_{1})\nonumber
\end{eqnarray}

\noindent
where $a_{i},(i=1,\,2)$ are the spinor amplitudes.
\par

In the atomic interpretation of the two-level system, $\Delta$ is 
the interlevel transition amplitude
and $\omega_{0}$ the level spacing.
\par
In the Landau-Zener (L-Z) regime, i.e. close to the level crossing,
assumed at $t=0$,
 one has:
$\omega_{0} =  \Omega \, t = B_{0} \Omega_{0} t$.
The asymptotics of the solution, when $\omega_{0}$ is linear in $t$, is a 
textbook topic; it  can be readily obtained by WKB methods.
Let us write the equation for the amplitude
 $a_{2}(t)$,  in the form:
\begin{eqnarray}
\label{A2}
 {d^2 \over dt^2} a_{2}(t)&=&-W^{2}(t) \cdot  a_{2}(t)\\
W(t)&=&{{\Omega t} \over 2} \sqrt{  1 + ({\Delta \over { \Omega  t}})^2 - 
 {2i \over{ t^2 \Omega}}}\nonumber
\end{eqnarray}
From the large time behavior of  equation \ref{A2}, 
we get the semiclassical momentum
$$p(t)= {{\Omega t} \over 2} + { \nu \over  t}\,\,
\hspace{.5 in} \,\,
 \nu= { \Delta^{2} \over { 4 \Omega }}$$ 
\par
It is  natural to choose the following scattering 
states:
$exp\{ \pm {{i \Omega t^2} / 4}\} \cdot (t)^{\pm i \nu}$.\\
The scattering matrix $S$ is obtained from the Weber functions,
who solve  Eq. \ref{A2}
 (see, e.g., \cite{PS}, and references therein).
The procedure is straightforward once
one realizes that  eq. \ref{A2} is  
the Schr\"odinger equation for an inverted
harmonic oscillator in $D=1$.
\par
The $S$ matrix identifies a rotation.
The branching properties of the solution, characterized by the 
single parameter $\nu$, 
uniquely determine $S$ and the rotation. 
 Since  $\nu$ is a constant in our system, 
each crossing  produces an identical rotation.
\par
Let us label $S$ with  $\theta,\,\phi$:
\begin{eqnarray}
\label{S}
S_{1,1}&=& cos(\theta)\,=\,exp\{- \pi \nu\}\nonumber\\
S_{1,2}&=& i \,sin(\theta) \cdot exp\{i \phi\}\,=
\,\sqrt{{ 2 \pi \over \nu}} \cdot exp\{ - (\pi \nu) /2 - 
(i \pi)/4\} \cdot {1 \over \Gamma(i \nu)}\nonumber\\
S_{2,1}&=& i \, sin(\theta) \cdot exp\{-i \phi\}\nonumber\\
S_{2,2}&=& S_{1,1}\nonumber
\end{eqnarray}
where $\Gamma(x)$ is the Gamma function. 
Clearly $\theta$ acts on the population $Z$, and  $\phi$ on the coherences
$X,\,Y$.
\par   
One finds, for the ratio of ``in'' and ``out'' populations, 
$(T\, = \,{Z(+ \infty) / Z(- \infty)})$: 
$T\, =\,T(\nu)\,=\,2 \cdot exp\{-2  \pi  \nu\} -1$. 

\par
So far, we discussed the Schr\"odinger equation.
Here  we solve the corresponding master equation problem.
Notice that, since in 
going from the spin variables  $a_{1,2}$  to the vector variables 
$v_{i}$ of the density matrix:
$v_{i}\,= a^{*}_{l} \sigma_{i}^{l,k} a_{k}\,\,(i=1,2,3),\,\,(l,k\,=\,1,2)$,  
one performs a nonlinear transformation, the new problem is not a 
straightforward 
translation of the old one.
\par
Rather than satisfying a second-order 
order equation as in Eq. \ref{A2}, each vector component satisfies a 
third-order equation, so that the connection with the Schr\"odinger
equation is lost.
\par
In spite of that, in the representation of frequencies $\omega$,  
the population ${\tilde Z}(\omega)$ satisfies  a Schr\"odinger equation 
for an (inverted) harmonic oscillator in D=2,  with a 
centrifugal term  corresponding to an ``angular momentum'' $m = \pm 1$.
\par
One can  develop an $S$ matrix formalism in this
representation, and  extract the L-Z rotation from 
the branching properties of the solution at $\omega \,=\,0$:
indeed at the level crossing  the frequency of the motion inverts its sign.
It should be further pointed out that, as long as $Z$ is quadratic
in the spinor amplitudes, its branching behavior is accordingly modified 
with respect to the spinor case.
In  Appendix II we give the main steps of the solution.
\par 
Before going to  numerical results, let us summarize the qualitative 
features of the motion:\\
a) in the ``normal'' regime the representative undergoes 
uniaxial rotations 
(precession with time-dependent frequency having an  average value of
$\pm 2 B_{0}/\pi $);
b) in the L-Z regime a discrete rotation, involving the population as well 
as the coherences, occurs;
c) in the next ``normal'' regime
the bias
  $\omega_{0}(t)$ has opposite 
sign with respect to regime (a), so that 
the sign of precession is inverted in the L-Z region.
\par
Let us consider  the following  3 situations:
$\,\,T \approx 1,\, 0, -1$. 
\par
 In the first case we have $(\nu \ll 1)$, so that, after crossing,  
an initially ``magnetized''  state 
is preserved, apart from a small correction; in the second case, 
where the transition probability equals $1/2$,  
it is turned into a ``fully unmagnetized'' state; 
in the third case $( \nu \gg 1)$, magnetization reversal occurs.
\par

Fig.\,1  is the three-dimensional plot of the representative ${\bf v}$. 
The interlevel coupling   $\Delta$ is of the order
of the frequency $\Omega_{0}$. 
\par
The L-Z crossings  
can be found in the regions where  the trajectory  
inverts its path.
Furthermore, at the crossings the vector 
 ${\bf v}$ undergoes a discrete rotation
 $\{\theta,\,\phi\}$.
The vertical angle  $\theta$ is determined by 
the  transition
probability between the two levels, which   
goes as $exp\{- 2 \pi \cdot \nu\}$:
here, where we started with $Z\,=\,-1$, one can see that  
$|Z|$ is reduced accordingly at each 
crossing. 

\par 
In Fig. 2, $Z(t)$ is plotted 
over 
10 cycles of the external field. 
The fast oscillations  correspond to the ``normal'' regime,
where $Z$ is on average  constant; the ``normal'' regimes are 
separated by kinks, 
produced by the crossings.
 
\par
Here we have $\theta\,\approx\, \pi/3$, as three kinks are needed to 
reach the complete
inversion of $Z$. 
The population $Z$
(as well as the  coherences $X,\,Y$)  undergoes large scale oscillations, 
following  the bias frequency.
A planar  $(X,\,Y)$ plot would  exhibit similar 
oscillations 
associated with $\phi$.
\par
One could argue that a zero-average oscillating bias 
$B_{z}\,=\,\omega_{0}(t)$ must generate a zero-average $Z(t)$.
On the  contrary,  ``single spin magnetization'' is possible, 
with hamiltonian
dynamics.
Indeed, by operating on the ``normal'' evolution through
$\Omega_{0}$, we obtained trajectories where $Z(t)$ keeps a fixed sign.
\par
This can be understood as follows.
Although the L-Z rotations  always act additively on ${\bf v}$,  
if  the 
interposed ``normal''
evolution inverts the planar components, the sequence of the $\theta$ rotations
acts on $Z(t)$ alternatingly rather than additively.
\par
Under these conditions if, e.g.,
 one starts with $Z(0)\,=\, -1$, and $\theta\,=\, \pi/3$, at the first
crossing  $Z$ is raised at $Z \approx 0.3$, but at the next crossing,
 it turns back to $Z\,=\,-1$.
An example is shown in Figs. 3, 4, 5, referring to 
$Z(t), \, {\bf v}(t), \, Z(B)$
respectively.
\par

In this non-adiabatic regime in general we found  strong 
irreversibility; the ``symmetry-breaking'' solution instead is almost
perfectly reversible.
This is made clear in Fig. 5, where practically no hysteretic effects
are found. 
 
\par

When the transition  probability equals ${1 \over 2}$, 
the population, when started at $Z\,=\,-1$,
 turns into a perfectly balanced superposition of the two
levels $(Z\,=\,0)$ and the vertical angle is: $\theta\,=\, \pi/2$.
The ``in-phase'' evolution follows  the sequence 
$Z\,=\,-1\,;\,0\,;\,1\,;\,0\,;-1...$, the ``out-of-phase''
evolution is given by $Z\,=\,-1\,;\,0\,;\,-1\,;\,0...$.
The latter case again
generates a completely reversible process.
\par 
At much  larger values of the transition probability 
the magnetization aligns  to the bias at each step.
The field evolves slowly with respect to the two-level
system: we are in the adiabatic regime,
and we obtain  very small deviations from reversibility
in the plots $Z\,=\,Z(B)$ (not shown here).
\par
 
\section{Irreversible dynamics: homogeneous case}

The simplest non-hermitean generalization of the previous
model includes population damping and decoherence.
This leads to an homogeneous equation for ${\bf v}$:

\begin{eqnarray}
\label{ME2}
{\dot X} \,+\, \gamma_{1} \, X &=& - \omega_{0} \, Y \nonumber\\ 
{\dot Y} \,+\, \gamma_{2} \, Y &=& \omega_{0} \, X \,-\, \Delta \, Z\\
{\dot Z} \,+\, \gamma_{3} \, Z &=& \Delta \, Y \nonumber
\end{eqnarray}

\par
In compact form, we have  $ {d \over dt} {\bf v} = {\hat B} \cdot {\bf v}$.
Notice that the isotropic case $\gamma_{1}=\gamma_{2}=\gamma_{3}$,
apart from a Poincar\'e 
sphere 
with decaying radius, is identical to the unitary case.
\par
We thus consider the general case.
The large time asymptotics of the solution of Eq.\ref{ME2}  can be 
determined from
the eigenvalues $p_{j}(t)\,$ of the operator ${\hat B}(t)$ \cite{FED}.
\par
We limit ourselves to few preliminary results,  for the sake of simplicity 
in the uniaxial case:
$\gamma_{1}\,=\,\gamma_{2}\,=\, \gamma_{r} ;\,\,\gamma_{3}\,=\, \gamma$.
\par
The equation for the right eigenvectors ${\bf E}_{j}(t)$  reads:
\begin{equation}
 {\hat B}(t) \cdot {\bf E}_{j}(t)\,=\,p_{j}(t) \cdot{\bf E}_{j}(t)\nonumber
\end{equation}
It can be shown \cite{FED} that the large time asymptotics is determined by 
the following fundamental solutions:
\begin{equation}
{\bf v}_{j}(t) \approx  
 exp\{i \cdot \int_{t_{in}}^{t} p_{j}(u)\cdot du\} \cdot {\bf E}_{j}(t).\\
\end{equation}
\par
Since here we have $j=1,2,3$, the eigenvalues can be analytically computed.

It is convenient to shift 
the variable $\,\,p\,\,$:$\,\,p \to y = p +
 \gamma_{r} + \delta;\,\, [\delta= (\gamma  - \gamma_{r})/ 3]$.
The eigenvalues are then:
\begin{eqnarray}
\label{EIGEN}
y_{1}&=&u+v\nonumber\\
y_{2,3}&=& -{1 \over2} [(u+v) \pm i  \sqrt{3}  (u-v)]
\end{eqnarray}
where $u^3\,v^3$ are the  roots of the equation:
\begin{eqnarray}
\label{EQU}
t^{2}&+& f  t - g^{3}=0\\
f&=&2 \delta [ (\Omega t)^2 + \delta^{2} -{1 \over 2}  \Delta^{2}]\nonumber\\
g&=&{1 \over 3} [ (\Omega t)^{2} + \Delta^{2} -3 \delta^{2}]\nonumber
\end{eqnarray}
In the range of parameters we are interested in, 
both roots of eq. \ref{EQU} are real-valued.
\par
We  keep  terms  up to the order $O(1 / t^{2})$ in the expansion 
in powers of $t$, and get:
\begin{eqnarray}
\label{EIGEN2}
p_{1}(t)& \approx & - \gamma - 2\, \gamma_{c} {1 \over (\Omega  t)^2}\nonumber\\
p_{2,3}(t)& \approx & - \gamma_{r} + \gamma_{c} {1 \over (\Omega  t)^2}
\pm i  [\Omega \, t +  {{2 \nu} \over t}]\\
\gamma_{c}&=& \delta  (\delta^{2} - {1 \over 2}  \Delta^{2})\nonumber
\end{eqnarray}
\par
The first eigenvalue gives the correction to the damping $\gamma$,
while  $p_{2,3}$  in their real part, give the correction to 
 $\gamma_{r}$.
It is worth noticing that these corrections 
act 
on  $\gamma\,$ and$\,\gamma_{r}$ with opposite signs: in other words, 
decoherence is slowed down at the
expense of damping, and viceversa, according with the sign of 
$\gamma_{c}$.
\par  
The imaginary parts of $p_{2,3}(t)$, related with the oscillating behavior 
and with the branching 
properties  around $t\,=\,0$, are not influenced by 
decoherence and damping.
One verifies that, apart from a factor of $2$, which is expected
in going from the spinor amplitude to the vector amplitude, the
result coincides with the semiclassical momentum $p(t)$, given
after eq. \ref{A2}.

\par
In the example of Fig. 6 we assume  pure decoherence 
 in the adiabatic regime:
the population $Z$  follows the bias.
Notice that indeed $Z$ is damped
as expected from  the  correction
to $\gamma$ computed previously(see eq. \ref{EIGEN2}, 
with $\gamma_{c}\, >\,0$).
\par
The values of 
$X$ and $Y$, between crossings, sustain an oscillating $Z$ 
with an $\langle Z \rangle$  sensitively different from zero.
\par
The symmetry breaking solution, found in the previous section,
survives to decoherence, as shown in Fig. 7. 
Here we have the  conditions of Fig. 3,
but the  decoherence parameters of Fig. 6.
One can realize a ``magnetized'' single spin,
by means of an oscillating bias.
 Here the magnetization is in the direction of the bias
field; the situation discussed in ref. \cite{FMO} is rather
different: there the magnetization is orthogonal to the bias
and $Z$ relaxes toward the ground state.
\par
When $\gamma_{c} \,<\,0$, in the adiabatic regime,
the dominant effect is that the population decay is enhanced 
and   $Z(t)$ rapidly evolves towards zero.
\par

\section{Inhomogeneous case: relaxation from upper to lower level}

The incoherent processes 
in a two-level system include 
 internal transitions (i.e. interlevel transitions
mediated by the surroundings).
\par
Here we analize the effects of incoherent
relaxation from the upper to the lower level.
\par
At the L-Z crossings 
 the ground state
flips between the two levels, and the inhomogeneous
term must change its sign accordingly; we are then led to the inhomogeneous
equation:

\begin{eqnarray}
\label{ME3}
{\dot X} &+& \gamma_{1} \, X = - \omega_{0} \, Y\nonumber\\ 
{\dot Y} &+& \gamma_{2} \, Y = \omega_{0} \, X \,-\, \Delta \, Z\\
{\dot Z} &+& \gamma_{3} \, (Z + {\omega_{0} \over |\omega_{0}|}) = \Delta \, Y\nonumber
\end{eqnarray}

If one were to choose a sign for the inhomogeneous term 
in Eq. \ref{ME3}, one would describe a 
situation in which the system would be forced to
decay towards a given level. 
This is appropriate only provided that $\omega_{0}$ has a fixed sign.
\par
The calculations presented in Fig. 8, 9 refer to the adiabatic, 
strongly damped regime. 
The population $Z$ starts switching
from a ``magnetized'' state to the opposite one, exactly when 
the external field
changes its sign.
\par
The escape from the instability is markedly different
from the relaxation towards the stable ground state (see Fig. 9), but this
is not the origin of
the hysteretic behavior shown in Fig. 8.

\par
In the approach towards a fully ``magnetized'' state,
the coupling between $Z$ and the coherences $X,\,Y$
plays a relevant role (see the oscillanting branches 
in the hysteresis cycle).
When the bias retraces its path backwards, the system is already in a
fully ``magnetized'' state $Z\,=\, \pm 1$, $X=Y=0$.
Only the onset of the instability, as the bias changes its sign,
is able to put the system in motion again.
\par
In conclusion,
we have two regimes (incoherent
$X,\,Y \approx 0$ and coherent $X,\,Y \ne 0$ 
respectively) both  compatible with the same value of the bias.
\par
So far we discussed a situation where the period $ 2 \pi /\Omega_{0}$
is much larger than the decay times of the two-level
system, so that the external field can be considered as quasi-static.
As the decay time gets longer, the system does not relax
fast enough towards the ``fully magnetized'' state:  an hysteretic behavior
 survives, as shown in Fig. 10.

\par
We conclude this section with a last example: we added the
relaxation towards the ground state to the system discussed in Fig. 7;
 the frequency $\Omega_{0}$ is comparable with $\Delta$. 
\par
In Fig. 11, 
the fastly oscillating portions refer to the ``normal'' evolution,
and the kinks to the crossings.
Notice that the oscillating parts are attracted towards the ``fully
magnetized'' state, which is never reached.
The system shows an hysteretic limit cycle (see Fig. 12). 

\section{Conclusions}

We examined the time evolution of a two-level system in the presence of an 
oscillating external field,
both in the unitary case and with decoherence and dissipation.
We solved, in this context, the Landau-Zener problem in the density matrix 
formalism, and showed that,
in the frequency representation, the population ${\tilde Z}(\omega)$ 
satisfies a Schr\"odinger equation for an inverted
harmonic oscillator in D=2. 
\par
We indicated a procedure allowing to interpret the various dynamical regimes 
by means of few intuitive rules.
We found that it is possible to sustain
a magnetization parallel to the bias for the single spin.
\par
We became recently aware of related work \cite{FMO}, where instead  the spin
is magnetized in the direction orthogonal to the bias;
furthermore, while our result holds in the homogeneous case,
in Ref. \cite{FMO} relaxation towards a single level is
assumed.
\par
We have also shown that one can  explicitly solve the Landau-Zener 
problem in the non-unitary case,
by means of a suitable extension of WKB methods. 
We stress that in our approach the parameters of the equation
are completely arbitrary, so that no ``adiabatic elimination'' 
of fast variables is needed \cite{LELO}.
\par
We find that
the  corrections to $\gamma$ and $\gamma_{r}$ have opposite sign.
The branching properties of the solution 
are not modified with respect to the unitary case:
 the scattering matrix  is again a function 
of the single L-Z parameter $\nu$.
\par
When  decay from the upper to the lower level is taken
into account, we find  hysteretic behavior: this effect originates
in the coexistence of a fully magnetized state, where the coherences
$X,\,Y$ are strictly zero, and a partly magnetized state, with
nonzero coherences. 
This interpretation  holds in the overdamped case; at slower dampings 
 an hysteretic limit cycle survives, related with non-adiabatic effects.

\section{ Appendix 1. Inequalities following from complete positivity}

The conditions for
complete positivity of the evolution, when written  for matrix $A_{ij}$, are:
$$
0 \; \leq \; \gamma_{1} \; \leq \; \gamma_{2} \,+\, \gamma_{3}
\hspace {.5 in}
0 \; \leq \; \gamma_{2} \; \leq \; \gamma_{3} \,+\, \gamma_{1}
\hspace {.5 in}
0 \; \leq \; \gamma_{3} \; \leq \; \gamma_{1} \,+\, \gamma_{2}
\hspace {.5 in}
$$
\par
\noindent
together with:

$$
4 \, (\gamma^{2} \,+\, \frac{1}{4}\, C_{1}^{2}) \; \leq \;
\gamma_{1}^{2} \,-\, ( \gamma_{2} \,-\, \gamma_{3})^{2} 
$$
$$
4 \, (\beta^{2} \,+\, \frac{1}{4}\, C_{2}^{2}) \; \leq \;
\gamma_{2}^{2} \,-\, ( \gamma_{3} \,-\, \gamma_{1})^{2} 
$$
$$
4 \, (\alpha^{2} \,+\, \frac{1}{4}\, C_{3}^{2}) \; \leq \;
\gamma_{3}^{2} \,-\, ( \gamma_{1} \,-\, \gamma_{2})^{2} 
$$
$$
4 \, \beta \, (\alpha \gamma \,-\, \frac{1}{4}\, C_{3} \, C_{1}) \,-\,
2 \, C_{2} \, ( \frac{1}{2}\,\alpha\, C_{1}
 \,+\, \frac{1}{2}\, \gamma \, C_{3}) \; \geq
\; (\gamma_{1} \,+\, \gamma_{3} \,-\, \gamma_{2} )(
\beta^{2} \,+\, \frac{1}{4} \, C_{2}^{2}) \,+\,
$$

$$
\,
(\gamma_{2} \,+\, \gamma_{1} \,-\, \gamma_{3})
[ \alpha^{2} \,+\, \frac{1}{4}\, C_{3}^{2} \,-\,
\frac{1}{4}\, \gamma_{3}^{2} \,-\, \frac{1}{4}\, ( \gamma_{2} \,-\, \gamma_{1}
)^{2} ] \,+\, 
(\gamma_{3} \,+\, \gamma_{2} \,-\, \gamma_{1})( \gamma^{2} \,+\, 
\frac{1}{4}\, C_{1}^{2})
$$
In our problem we further must take into account $\omega_{0}(t)$,
which defines the scale of times, and  $\Delta$.
Thus $\Delta$ is  the single relevant parameter of the model.

\vspace {1 cm}

\par
\section{Appendix II. Solution of the Landau-Zener problem from the 
master equation}

In this Appendix we summarize the main points of the solution of the L-Z
problem, as obtained by starting from the master equation; we 
omit those details that can be easily implemented from the standard
solution. 
\par
In the non-dissipative case the Landau-Zener problem is written in the form:

\begin{eqnarray}
\label{LZunitary1}
{\dot X} &=& -(\Omega  t) \cdot Y\nonumber\\
{\dot Y} &=& (\Omega   t) \cdot X - \Delta \cdot Z\\
{\dot Z} &=& \Delta \cdot Y\nonumber
\end{eqnarray}

The variable $Y$ can be eliminated from the third equation, and one
obtains:
\begin{eqnarray}
\label{LZunitary2}
{\ddot Z} &+&  \Delta^{2} \cdot Z = \Delta \Omega  t \cdot X\\
{\dot X} &=& - { \Omega \over { \Delta}}  t \cdot {\dot Z}\nonumber
\end{eqnarray}
\par
Let us define the new variable $\phi (t)\,=\,X(t)+{\Omega \over { \Delta}} t \cdot Z(t)$;
it is readily verified, by inspecting  the second equation 
in system \ref{LZunitary2}, that from $\phi (t)$ one can
recover the entire solution: $Z(t)\,=\,(\Delta / \Omega)
 \cdot {\dot \phi (t)};\,\,
X(t) \,=\, \phi(t) -t \cdot \phi (t)$.
\par
The Fourier transform $\tilde {\phi} ( \omega)\,=\, \int^{+ \infty}_{- \infty} exp\{- i  \omega  t\} \cdot \phi(t) \cdot dt$
solves the equation:
\begin{eqnarray}
\label{LZunitaryFT}
\hat{ H} \tilde{\phi}(\omega)&=&0\nonumber\\
\hat{H}= - { \omega ^{3} \over  \Omega^{2}} &+& ({  \Delta^{2} \over   
\Omega ^{2}} - { \partial ^2 \over {\partial  \omega^2}})  \omega
 - {\partial \over {\partial \omega}}\nonumber\\
\end{eqnarray}
Let us write $H$ as:
\begin{eqnarray}
\hat{H}&=& \hat{h} \cdot \omega\nonumber\\
\hat{h}&=&-{ \partial^2 \over { \partial \omega^2}} -{ 1 \over \omega}{\partial \over {\partial \omega}} +{1 \over \omega^2}
- ({ \omega \over \Omega})^2 +( {\Delta \over \Omega})^2 
\end{eqnarray}

We find that the function 
$\chi(\omega)\,=\, \omega \cdot \tilde{\phi}(\omega)$, which, 
apart from a constant factor,
is the Fourier transform of the population  $Z(t)\,:\,( \chi(\omega)\,=\, { \Omega \over { \Delta}} \cdot \tilde{Z}(\omega) ) $, 
satisfies the Whittaker equation \cite{GR}:
\begin{equation}
\label{Whittaker}
\hat{h} \chi(\omega)\,=\,0.\\
\end{equation}
\par
Notice that eq. \ref{Whittaker} has the form of a Schr\"odinger equation  for the inverted harmonic oscillator in D=2, with ``angular momentum'' $m\,=\, \pm 1$; (clearly in our case  $\omega$ varies over the whole axis).
\par
The equation can be written in the canonical form of the  hypergeometric 
confluent equation (HCE)  by letting:
$q= \omega^2\,;\,\chi(\omega)\,=\,
exp\{{{i  q} \over {2 \Omega}}\} \cdot (q)^{1/2} \cdot u(q)$.
\par
In terms of the new  variable  $q\,:\, q\,=\,(i \Omega)\cdot \xi$, 
one finally gets, for  the function $w(\xi)\,=\,u(q)$:
\begin{eqnarray}
\label{HCE}
\xi \cdot {  d^2 \over d \xi^2} w&+&(2- \xi) \cdot {d \over d \xi} w-
(1+i \nu) \cdot w =0\\
\nu = {\Delta^{2} \over {4 \Omega}} &.& \nonumber
\end{eqnarray}
\par
Two linearly independent solutions, respectively regular at infinity and around zero, are:
\begin{eqnarray}
\label{SOL}
\chi_{1}(\omega)&=&exp\{{{i \omega^2} \over {2 \Omega}}\} \cdot \omega \cdot \Psi(a,2,{{-i \omega^2} \over  \Omega})\nonumber\\
\chi_{2}(\omega)&=&exp\{{{i  \omega^2} \over {2 \Omega}}\} \cdot \omega \cdot \Phi(a,2,{{-i  \omega^2} \over  \Omega})\nonumber\\
a=1&+&i \nu\nonumber
\end{eqnarray}
where $\Phi,\,\Psi$ are the hypergeometric confluent functions of first and second type.
\par
In the L-Z  problem the large time region is characterized by high frequencies, since there  the level spacing
always dominates over the interlevel coupling.
Hence the solution $\chi_1(\omega)$, having a well-defined behavior 
at infinity, 
is the natural choice in our case. 
\par
We consider the behavior of the function $\Psi(a,c,z)$:
$$
\Psi(a,c,z) \approx (z)^{-a} \cdot [1 +O({1 \over z})],\,( z \gg 1),
$$
\noindent
and  recall that $\Psi$ satisfies the identity:
$$
\Psi(a,c,z)\,=\,(z)^{1-c} \cdot \Psi(a-c+1,2-c,z).
$$
\par
If one takes the former properties into account,  
and uses the identity:
${\tilde \phi}(\omega)\,=\,{\tilde \phi}^{*}(- \omega)$),
which follows from $\phi(t)$ being real, 
one obtains:
\begin{eqnarray}
\label{PHI}
{\tilde \phi}(\omega)&=& A\, exp\{{{i \omega^2} \over {2 \Omega}}\} \cdot 
{1 \over \omega^{2}} \cdot \Psi(i  \nu,0,{{-i  \omega^2} \over  \Omega})\nonumber\\&+&
A^{*}\,  exp\{{{-i \omega^{2}} \over {2 \Omega}}\} \cdot 
{1 \over \omega^2} \cdot \Psi(-i  \nu,0,{{i  \omega^2} \over  \Omega})\nonumber\\
\end{eqnarray}
\par
One  recovers the behavior of the population
 from ${\tilde Z}(\omega)=\,i\,{ \Delta \over \Omega}\, 
\omega \cdot {\tilde \phi}(\omega)$;
we have, as 
$\omega \to \infty ,{\tilde Z}(\omega) \approx 1/\omega$.
\par 
More precisely, apart from
constant factors:
\begin{equation}
\label{Z}
{\tilde Z}(\omega)\,=\,i {\Delta \over {\Omega \omega}} \cdot 
exp\{- \pi  \nu/2\} \cdot [ exp\{i  {\omega^2 \over {2  \Delta}}\} \cdot ({\omega \over \sqrt{\Omega}})^{-2 i  \nu}
+ exp\{-i  {\omega^2 \over {2 \Delta}}\} \cdot ({\omega \over \sqrt{\Omega}})^{2 i  \nu}]\\
\end{equation}
\par
It is easily verified that the correct transition probabilities 
can be extracted from the branching properties around $\omega\,=\,0$ 
of the solution ${\tilde Z}(\omega)$.
\par
We remind that  the standard treatment
refers to spinor amplitudes, while ${\tilde Z}(\omega)$ is quadratic
in such amplitudes: this explains the factor of $2$ in the exponent 
of $\omega$.
\par
The coherence ${\tilde X}(\omega)$, can be found from the identity: 

$$
 {\tilde X}(\omega)={\tilde \phi}(\omega)+
 {d \over {d \omega}}( \omega {\tilde \phi}(\omega)).
$$
\par
We add the large $\omega$ behavior of ${\tilde X}(\omega)$:
\begin{equation}
\label{X}
{\tilde X} (\omega) \approx {1 \over \Omega} (1 - 2 ({\Delta \over \omega})^2) 
exp\{- \pi \nu)/2\} [ exp\{i {\omega^2 \over {2 \alpha}}\} 
\cdot ({\omega \over \sqrt{\Omega}})^{-2 i \nu}
- exp\{-i  {\omega^{2} \over {2 \alpha}}\} 
 ({\omega \over \sqrt{ \Omega}})^{2 i \nu})]\nonumber
\end{equation}
\par
In summary, while the Schr\"odinger equation leads to the inverted harmonic 
oscillator
in D=1 (Weber equation) 
in the representation of times, 
the master equation leads, for ${\tilde Z}(\omega)$, 
to the D=2 inverted harmonic
oscillator.
Hence, in the representation of frequencies,
 it is  possible to formulate and solve the L-Z problem in terms of 
the $S$ matrix.
\par

\newpage

\begin{figure}
\begin{center}
\includegraphics[scale=0.9,angle=-90]{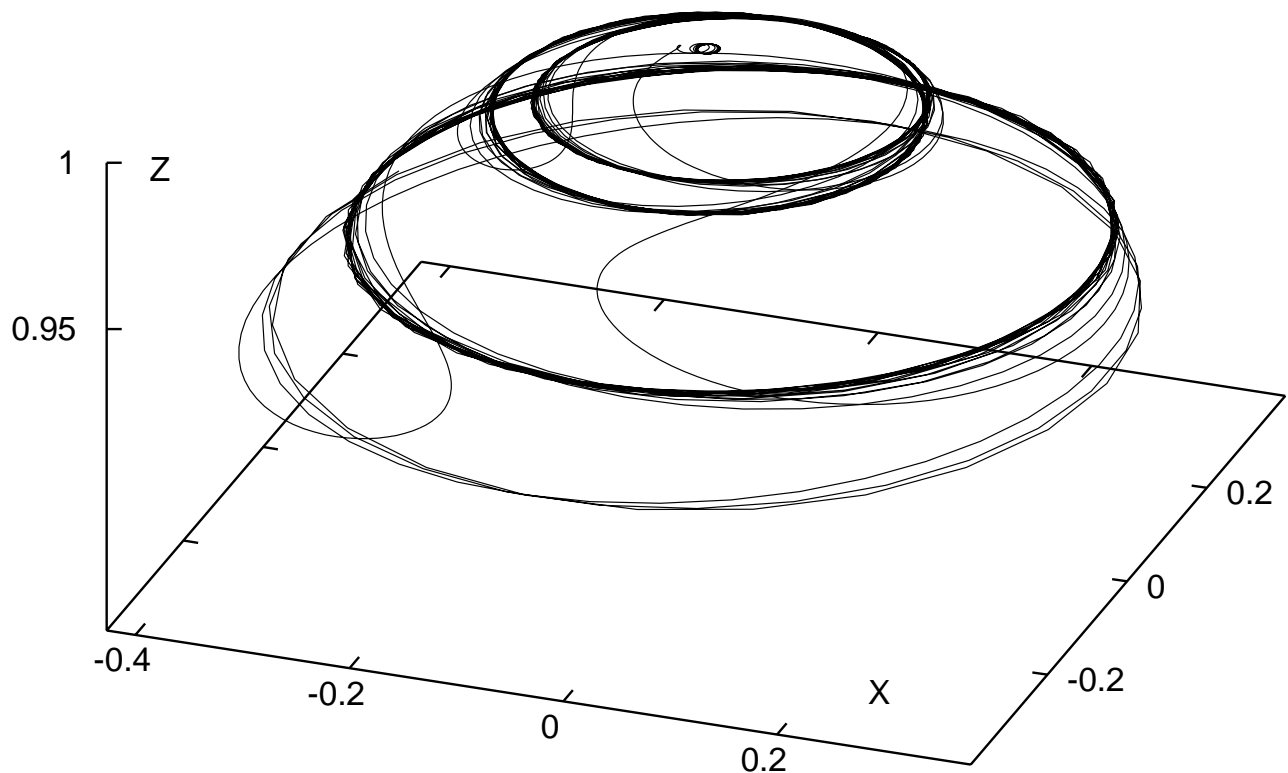}
\end{center}
\caption{ A typical 
trajectory ${\bf v}(t)$,$\,{\bf v}\,=\,X,Y,Z$, starting
at  $Z\,=\,1$. The ``normal'' evolution corresponds to the
fast uniaxial rotations: here four branches of this sort can be 
identified. At the crossings the trajectory inverts its path and
makes a transition to a new branch.
We have $\Delta\,=\,0.01$, $\Omega_{0} \,=\,0.02$;
here and in all other figures $B_{0}=1$.}
\end{figure}
\par
\begin{figure}
\begin{center}
\includegraphics[scale=0.7,angle=-90]{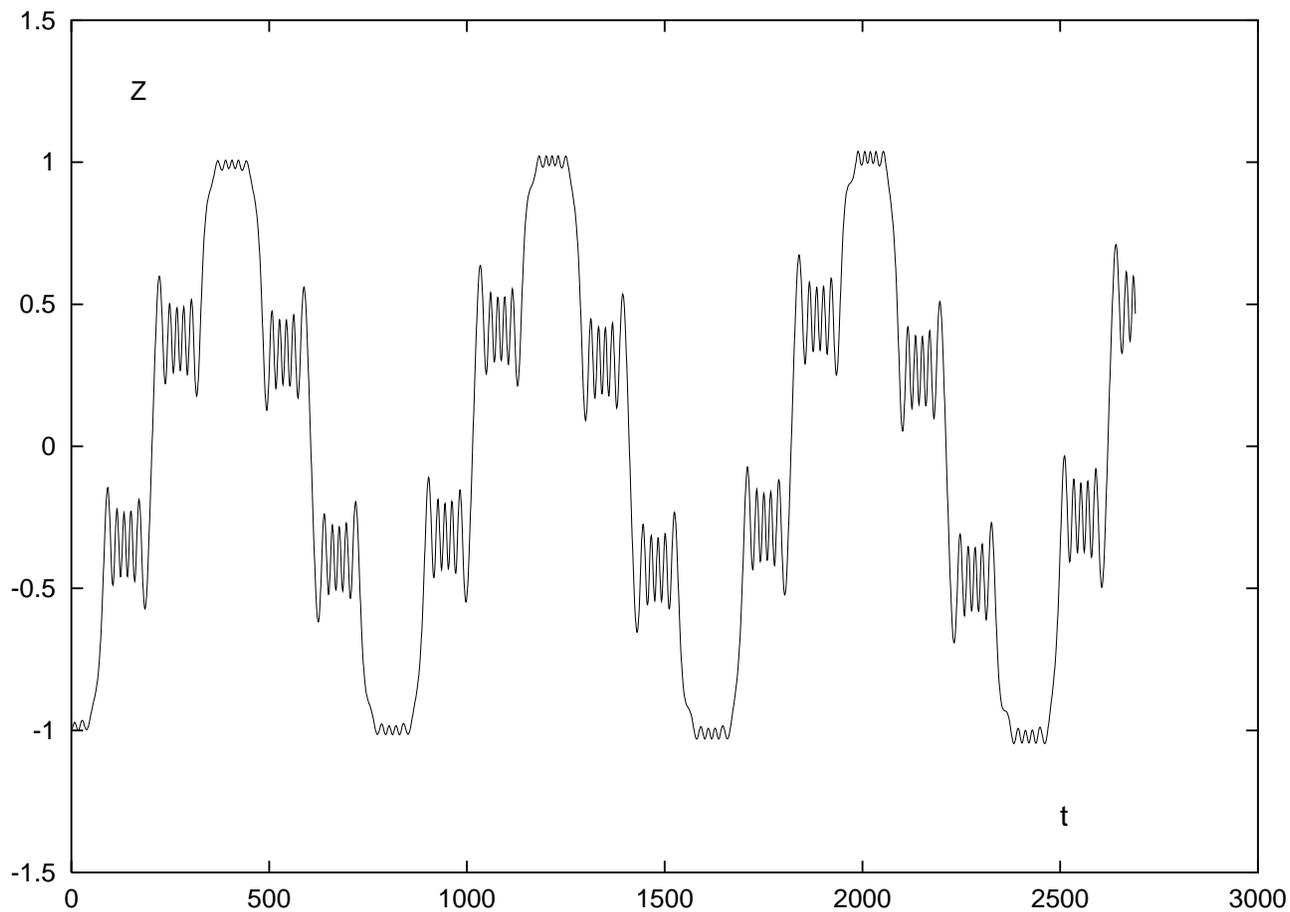}
\end{center}
\caption{ Time evolution of $Z$ over 10 cycles of the bias field.
Three kinks are needed in order to have a complete inversion of
the magnetization: hence $\theta\,=\, \pi/3$.
Here $\Delta\,=\,0.12$, $\Omega_{0}\,=\, 0.063$.
Here and in all the Z(t) plots the scale of time is in arbitrary units.}
\end{figure}
\par

\begin{figure}
\begin{center}
\includegraphics[scale=0.7,angle=-90]{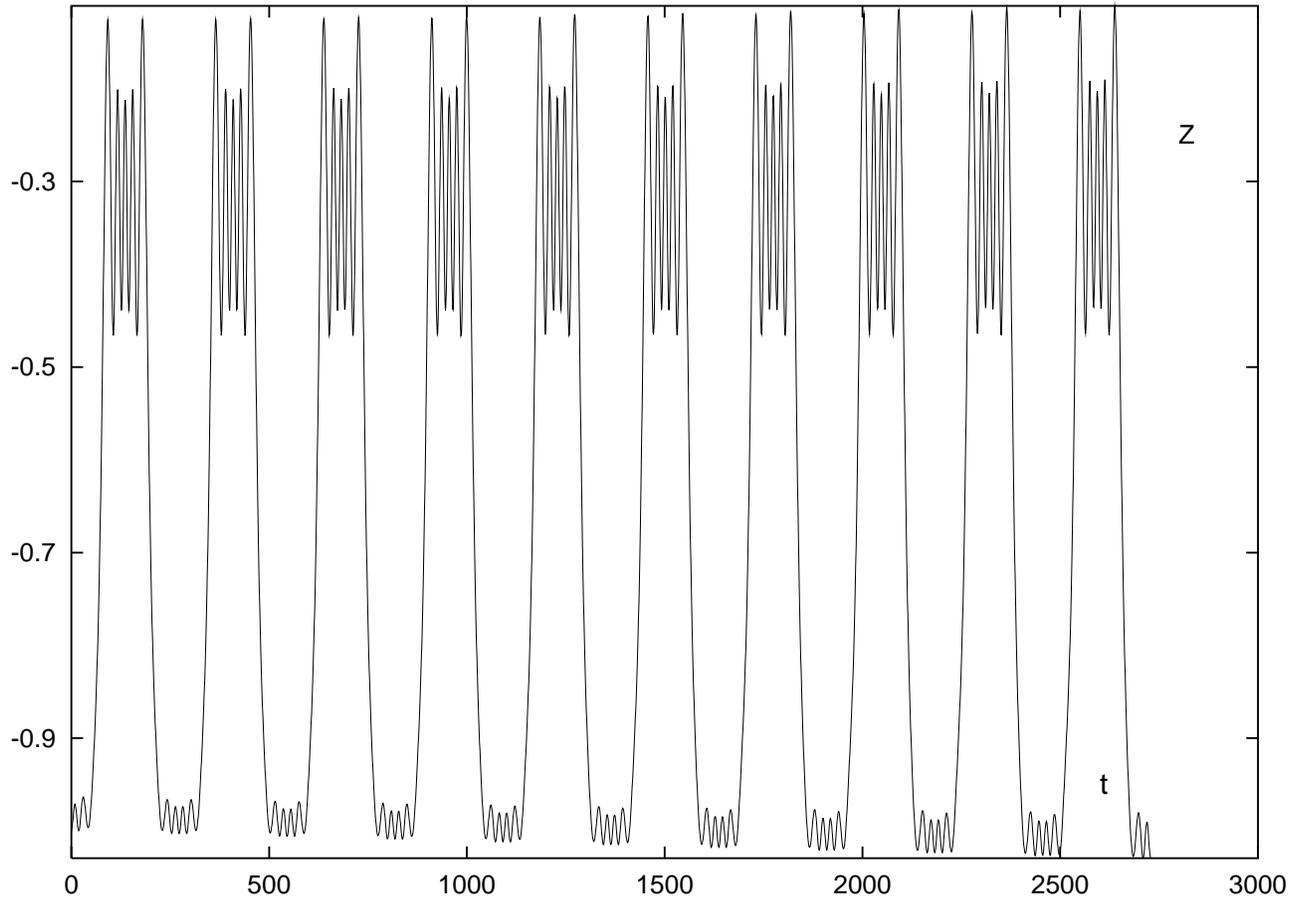}
\end{center}
\caption{ Time evolution of $Z$ over 10 cycles of the bias field,
with 
$\Delta\,=\,0.12$, $\Omega_{0} \,=\, 0.0682$.
With this choice, at each crossing
the vertical rotation changes its sign: the result is an 
explicit symmetry breaking: $<Z(t)> \,<\,0$.}  
\end{figure}
\par
\begin{figure}
\begin{center}
\includegraphics[scale=0.9,angle=-90]{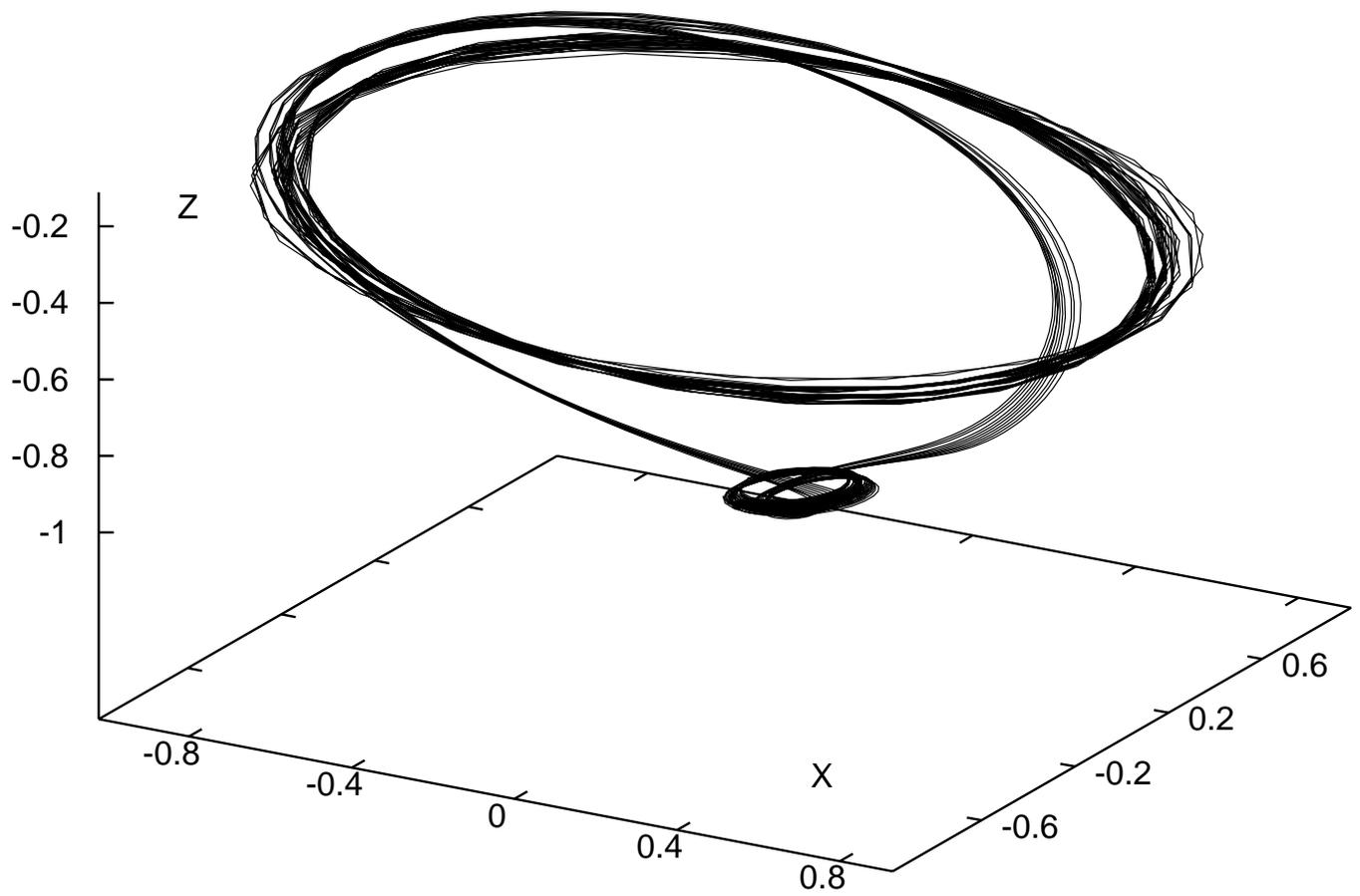}
\end{center}
\caption{ Trajectory {\bf v}(t), parameters as in Fig. 3.}
\end{figure}
\par

\begin{figure}
\begin{center}
\includegraphics[scale=0.7,angle=-90]{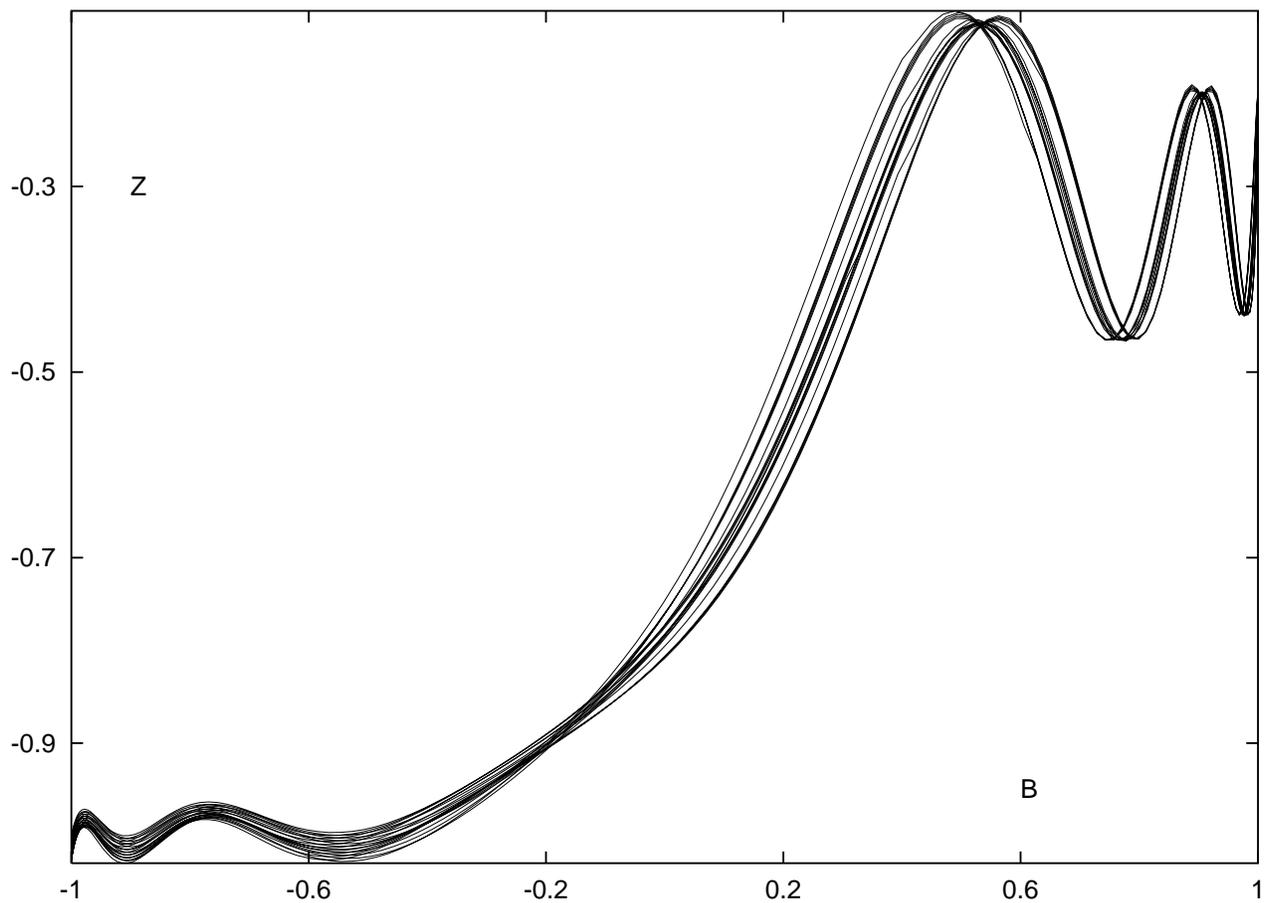}
\end{center}
\caption{ Plot of $Z\,=\,Z(B)$, from the solution of Fig. 3.
Notice that, in spite of being far from the adiabatic regime,
the system shows very slight deviations from complete reversibility.
These deviations are very sensitive when the system is tuned
out of the particular regime displayed in Figs. 3, 4, 5.}
\end{figure}
\par
\begin{figure}
\begin{center}
\includegraphics[scale=0.7,angle=-90]{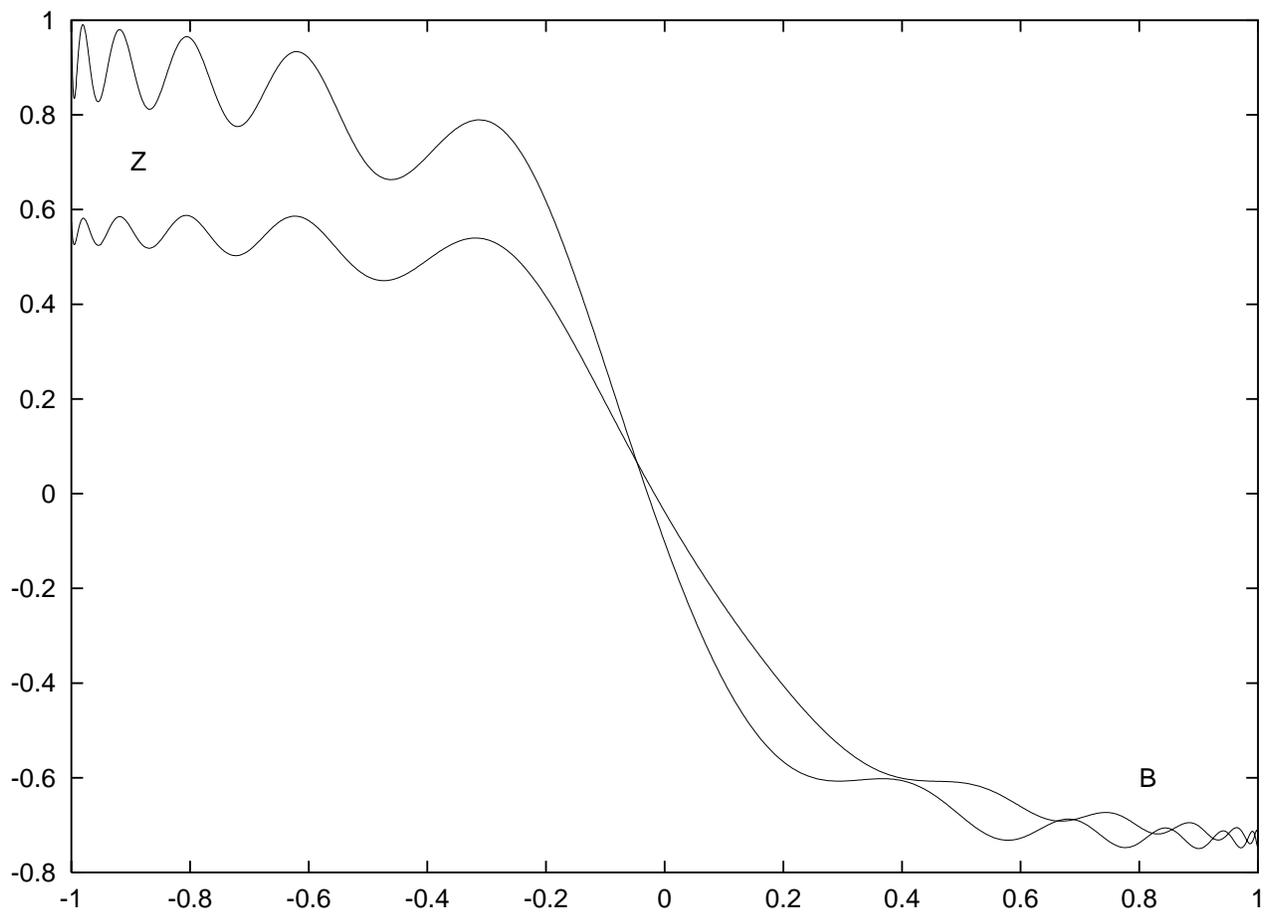}
\end{center}
\caption{ Plot of $Z(B)$ for a single cycle of the bias, with initial
condition $Z(t=0)=1$.The parameters are 
$\Delta\,=\,0.3$, $\Omega_{0} \,=\, 0.033$, 
$\gamma_{r}\,=\,0.01, \gamma\,=\,0$.}
\end{figure}
\par

\begin{figure}
\begin{center}
\includegraphics[scale=0.7,angle=-90]{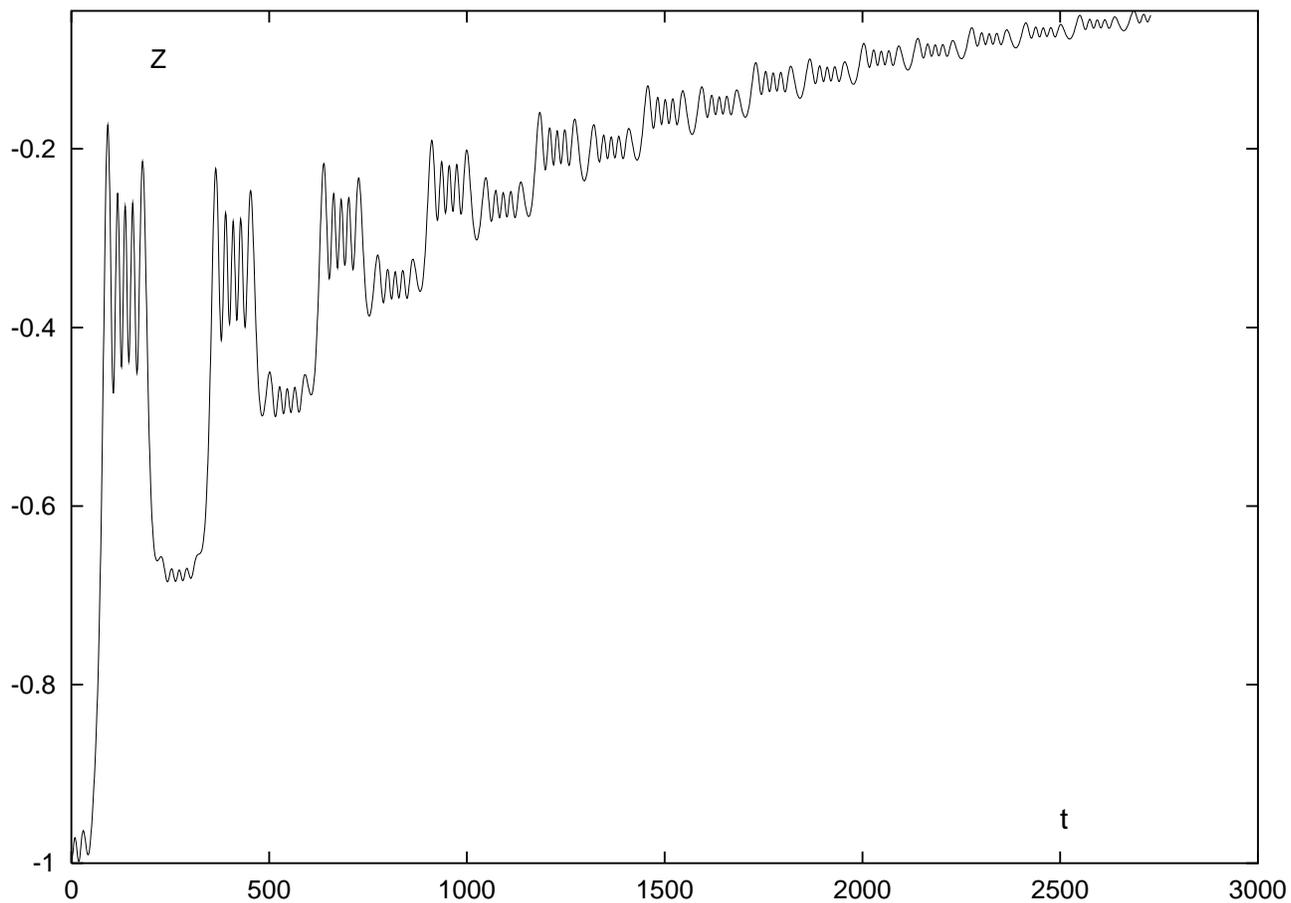}
\end{center}
\caption{ Survival of magnetization in the presence of decoherence:
plot of $Z(t)$ over ten periods of the bias, with
$\Delta\,=\,0.12$, $\Omega_{0} \,=\, 0.0682$,
$\gamma_{r}\,=\,0.01,\gamma\,=\,0$}
\end{figure}
\par

\begin{figure}
\begin{center}
\includegraphics[scale=0.7,angle=-90]{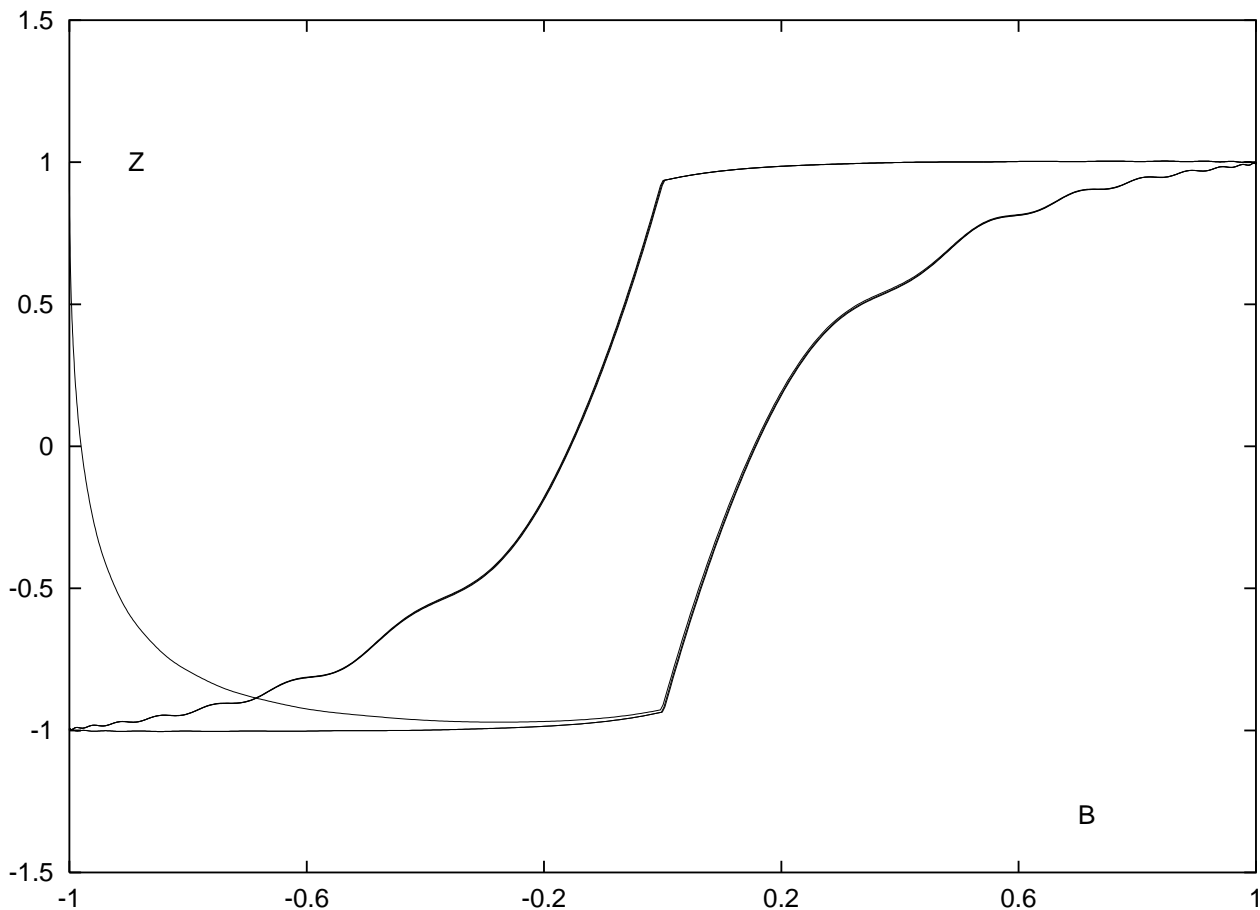}
\end{center}
\caption{ Approach to the hysteretic limit cycle in the strongly damped
case: $Z(B)$ with $\gamma_{r}\,=\,0.035,\gamma\,=\,0.07$ 
and 
$\Delta\,=\,0.05$, $\Omega_{0} \,=\, 0.02$.
 Notice that $Z$ is almost perfectly
constant in the  fully magnetized branches, where the coherences
$X,\,Y$ are zero. The magnetization reversal starts at
the zeros of the bias field.}
\end{figure}
\par

\begin{figure}
\begin{center}
\includegraphics[scale=0.7,angle=-90]{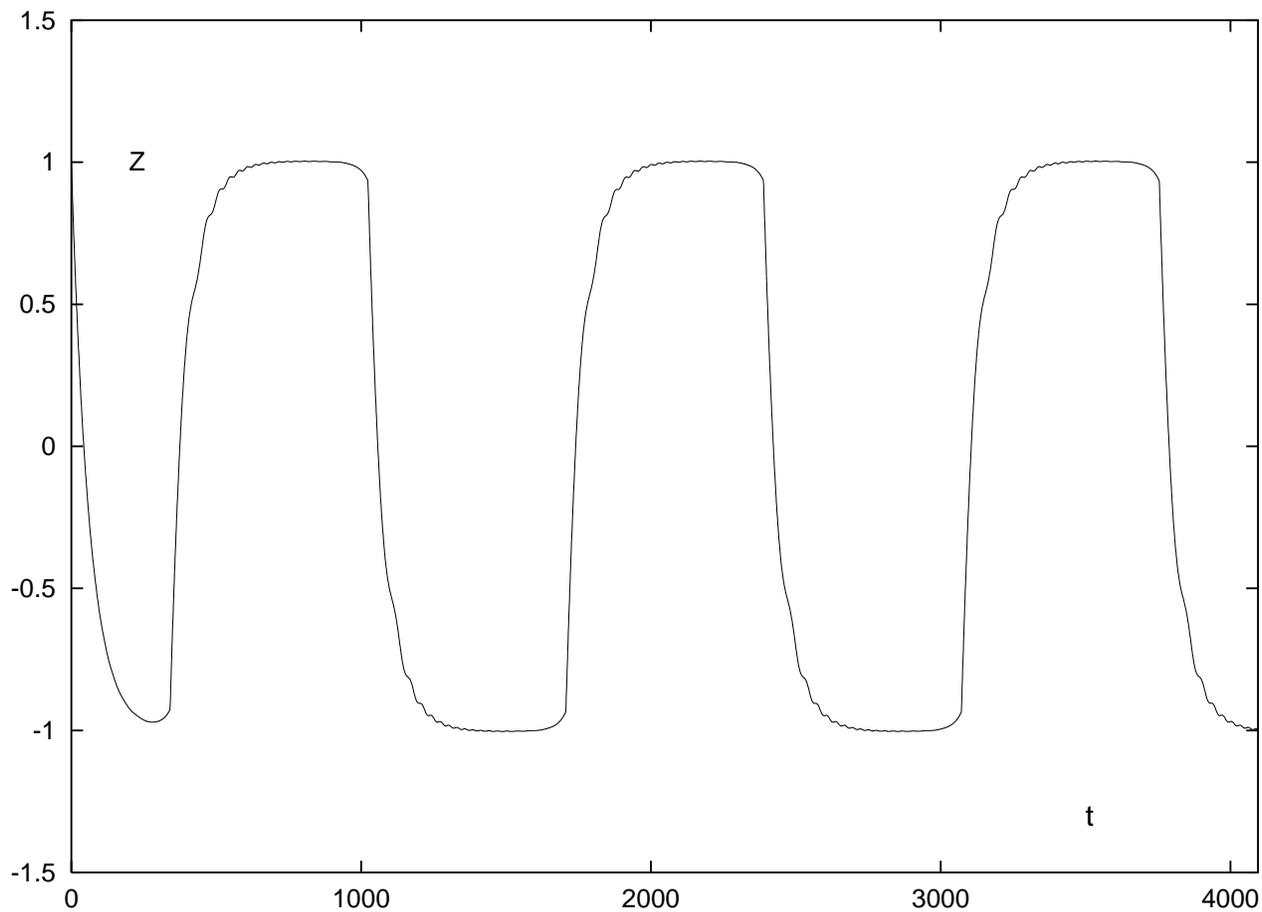}
\end{center}
\caption{ Plot of the 
magnetization $Z(t)$ in the same conditions of Fig.6a:
the cusps corresponding to the onsets of magnetization reversal
separate perfectly identical, but inverted, kinks.
The result is a sequence of pulses lacking specular symmetry.}
\end{figure}
\par
\begin{figure}
\begin{center}
\includegraphics[scale=0.7,angle=-90]{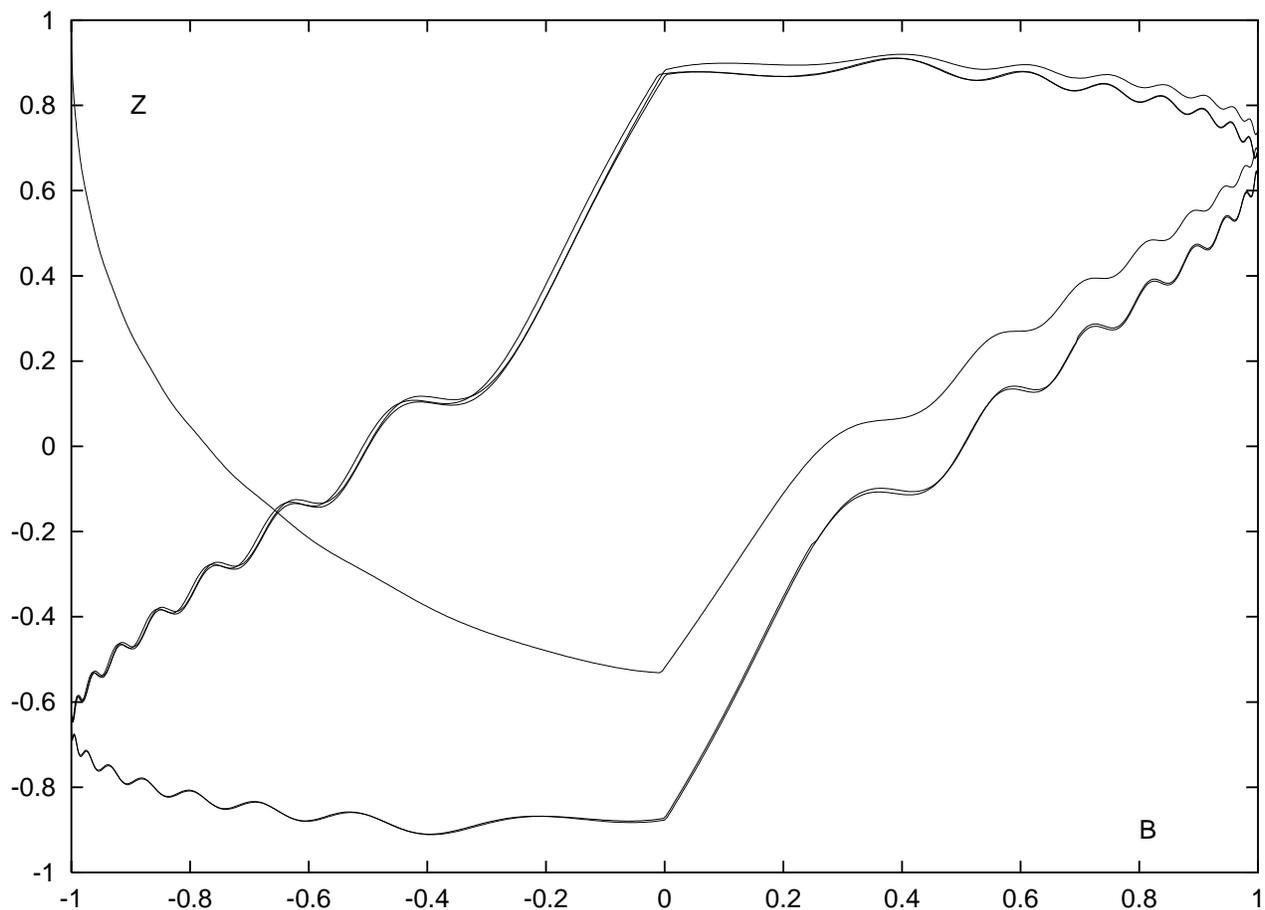}
\end{center}
\caption{ Same as in Fig.6a, but with smaller damping:
$\gamma_{r}\,=\,0.01,\,\gamma\,=\,0.02$. Here the complete magnetization
is never reached, because the
relaxation towards the ground state has a time scale comparable with
the period of the bias. An hysteresis is still obtained; the behavior of
$Z(t)$, not displayed here, is again a sequence of asymmetric pulses.}
\end{figure}
\par

\begin{figure}
\begin{center}
\includegraphics[scale=0.7,angle=-90]{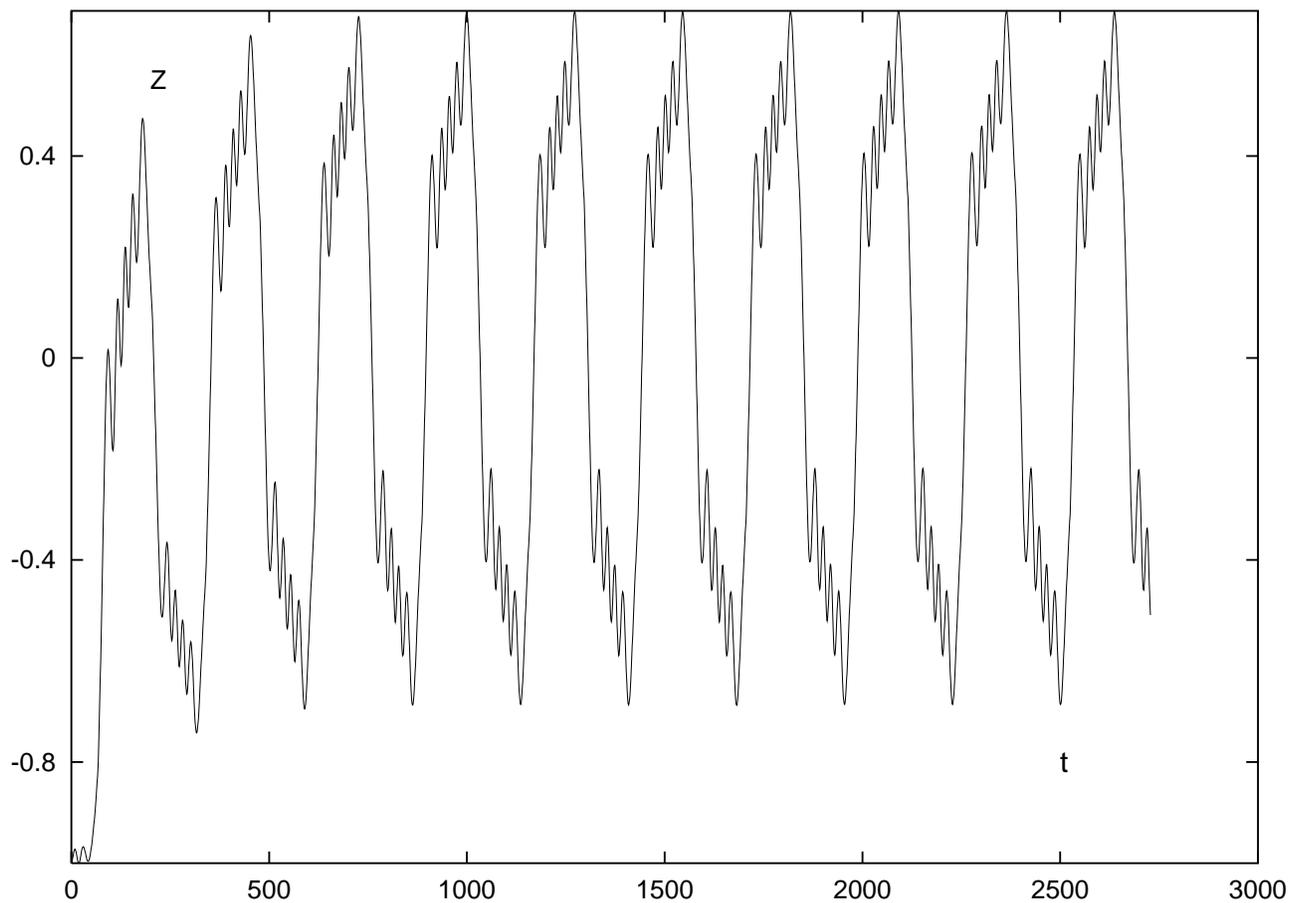}
\end{center}
\caption{ Plot of $Z(t)$ in the non-adiabatic regime; the parameters are
as in Fig.5, but now we added the relaxation towards ground state
with $\gamma\,=\,0.02$. It is instructive to compare this result with 
the hamiltonian case of Fig.3b, where we had the same values of 
$\Delta$ and $\Omega_{0}$: there $Z$ never changes its sign, here it
is attracted towards the lower level. 
Again each pulse is strongly asymmetric.}
\end{figure}
\par
\begin{figure}
\begin{center}
\includegraphics[scale=0.7,angle=-90]{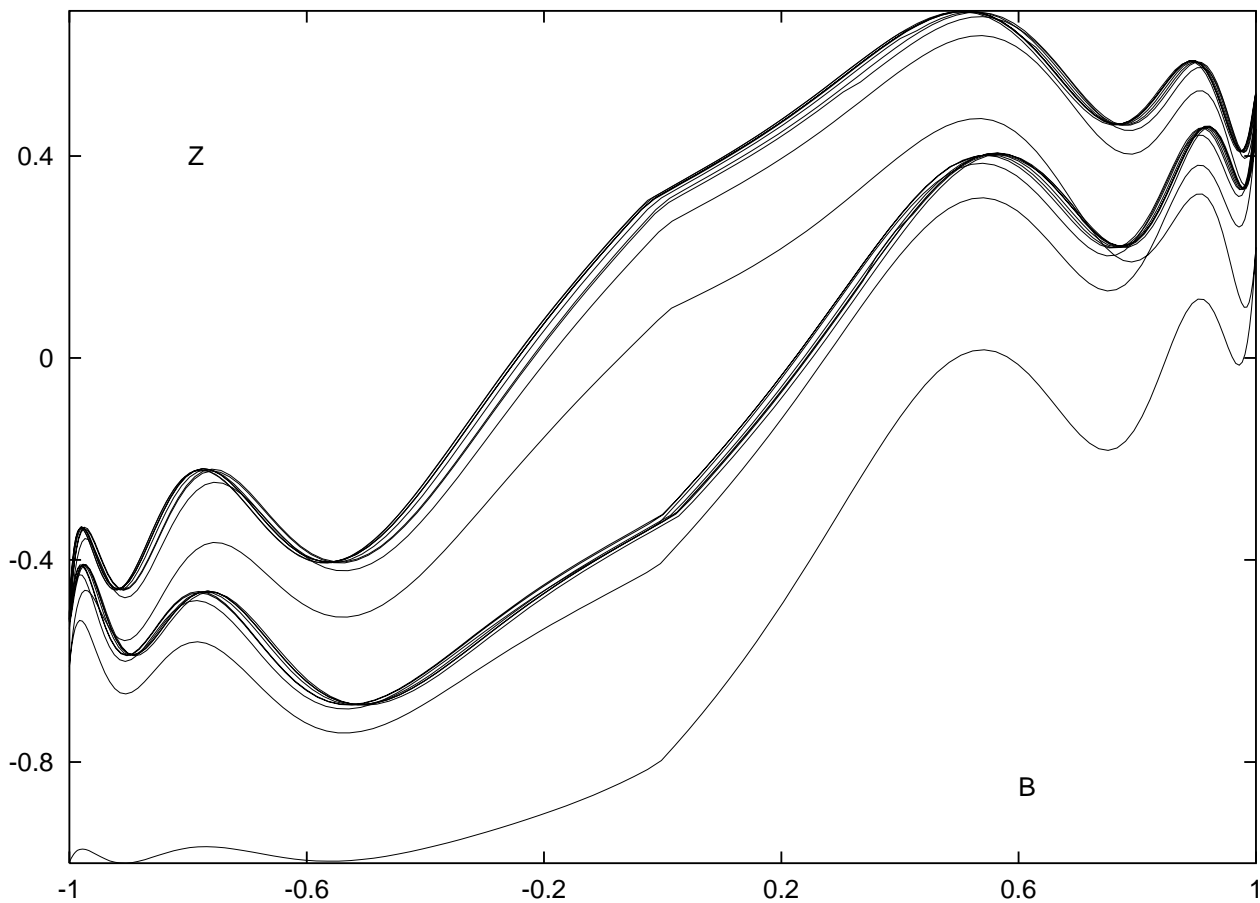}
\end{center}
\caption{ Approach to the hysteretic limit cycle of the solution 
discussed in Fig.8a: one can identify the cusps at the onset
of magnetization reversal.}
\end{figure}
\par

\section*{References}
\begin{thebibliography}{18}
\par
\bibitem{TB}I. Tupitsyn, B. Barbara in: 
 {\it Magnetoscience - From Molecules to Materials},
 Miller, Drillon Eds., Wiley VCH Verlag Gmbh (2000).
\par
\bibitem{DKH}V. V. Dobrovitski, 
M. I. Katsnelson, B. N. Harmon: {\it Phys. Rev. Lett.} {\bf 84}, 3458 (2000).
\par
\bibitem{DS}M. Dub\'e, P. C. E. Stamp: cond-mat/0102156
\par
\bibitem{CWMBB1}I. Chiorescu, W. Wernsdorfer, A. M\"uller, H. B\"ogge, B. Barbara:
{\it Phys. Rev. Lett.} {\bf 84}, 15, 3454 (2000).
\par
\bibitem{CWMBB2}I. Chiorescu, W. Wernsdorfer, A. M\"uller, H. B\"ogge, B. Barbara:
{\it J. Magn. Mat.} {\bf 221}, 1-2, 103 (2000).
\par
\bibitem{FMO}S. Flach, A. M. Miroshnichenko, A. A. Ovchinnikov: 
quant-ph/0110113.
\par
\bibitem{BBGW}V. G. Bagrov, J. C. A. Barata, D. M. Gitman, W. F. Wreszinski:
{\it J. Phys. } 
{\bf A35}, 175 (2002).
\par
\bibitem{BC}J. C. A. Barata, D. A. Cortez: quant-ph/0202110.
\par
\bibitem{LL}L. D. Landau, E. M. Lifshitz: {\it Quantum mechanics},
 Pergamon Press,
Oxford (1967).
\par
\bibitem{Z}C. Zener: {\it Proc. Roy. Soc. Lond.} {\bf A137}, 696 (1932).
\par
\bibitem{AL}R. Alicki and K. Lendi. {\it Quantum Dynamical Semigroups and
Applications}.
 Lect. notes in Phys. 286. Springer-Verlag (1987).
\par
\bibitem{RSW}M. B. Ruskai, S. Szarek, E. Werner: cond-mat/0101003;
 to appear in:
{\it Li. Alg. Appl.}
\par
\bibitem{K}K. Kraus, {\it Ann. Physics} {\bf 64}, 311 (1971).
\par
\bibitem{LELO}M. N. Leuenberger, D. Loss: {\it Phys. Rev.} {\bf B61},
 12200 (2000).
\par
\bibitem{FED}M. Fedoriuk: {\it Methodes asymptotiques pour les equations
 differentielles ordinaires lineaires}, Mir, Moscou, 1987.
\par
\bibitem{S} G. Strini: {\it Fortsch. der Phys}. {\bf 50}, 169 (2002); 
{\it Lecture Notes on Quantum Computing} (Unpublished, 2000).
\par
\bibitem{PS} V. L. Pokrovsky, N. A. Sinitsyn: cond-mat/0012303
\par
\bibitem{GR}I. S. Gradshtein, I. W. Rhyzik, 
{\it Table of Integrals, Series and Products.} (Academic Press, New York, 1994)\\
\par
\end{thebibliography}
\end{document}